\documentclass[twocolumn,english,aps,prd,nofootinbib]{revtex4}
\usepackage{lmodern}
\usepackage[T1]{fontenc}
\usepackage[latin9]{inputenc}
\usepackage{xcolor}
\usepackage{babel}
\usepackage{amsmath}
\usepackage{amssymb}
\usepackage{cancel}
\usepackage{esint}
\usepackage[unicode=true]
 {hyperref}

\makeatletter

\newcommand{\lyxmathsym}[1]{\ifmmode\begingroup\def\b@ld{bold}
  \text{\ifx\math@version\b@ld\bfseries\fi#1}\endgroup\else#1\fi}

\@ifundefined{textcolor}{}
{%
 \definecolor{BLACK}{gray}{0}
 \definecolor{WHITE}{gray}{1}
 \definecolor{RED}{rgb}{1,0,0}
 \definecolor{GREEN}{rgb}{0,1,0}
 \definecolor{BLUE}{rgb}{0,0,1}
 \definecolor{CYAN}{cmyk}{1,0,0,0}
 \definecolor{MAGENTA}{cmyk}{0,1,0,0}
 \definecolor{YELLOW}{cmyk}{0,0,1,0}
}

\usepackage{enumerate}

\makeatother

\begin{document}
\title{Emerging Newtonian potential in pure $R^{2}$ gravity on a de Sitter
background}
\author{Hoang Ky Nguyen$\,$}
\email[\ \ ]{hoang.nguyen@ubbcluj.ro}

\affiliation{Department of Physics, Babe\c{s}--Bolyai University, Cluj-Napoca
400084, Romania}
\date{August 8, 2023}
\begin{abstract}
\vskip2pt In $\,$Fortsch.$\,$Phys. \textbf{64}, 176 (2016), Alvarez-Gaume\emph{
et al }established that pure $\mathcal{R}^{2}$ theory propagates
\emph{massless} spin-2 graviton on a de Sitter (dS) background but
not on a locally flat background. We build on this insight to derive
a Newtonian limit for the theory. Unlike most previous works that
linearized the metric around a locally flat background, we explicitly
employ the dS background to start with. We directly solve the field
equation of the action $(2\kappa)^{-1}\int d^{4}x\sqrt{-g}\,\mathcal{R}^{2}$
coupled with the stress-energy tensor of normal matter in the form
$T_{\mu\nu}=Mc^{2}\,\delta(\vec{r})\,\delta_{\mu}^{0}\,\delta_{\nu}^{0}$.
We obtain the following Schwarzschild--de Sitter metric $ds^{2}=-\Bigl(1-\frac{\Lambda}{3}r^{2}-\frac{\kappa c^{2}}{48\pi\Lambda}\frac{M}{r}\Bigr)c^{2}dt^{2}+\Bigl(1-\frac{\Lambda}{3}r^{2}-\frac{\kappa c^{2}}{48\pi\Lambda}\frac{M}{r}\Bigr)^{-1}dr^{2}+r^{2}d\Omega^{2}$
which features a potential $V(r)=-\frac{\kappa c^{4}}{96\pi\Lambda}\frac{M}{r}$
with the correct Newtonian tail. The parameter $\Lambda$ plays a
dual role: (i) it sets the scalar curvature for the background dS
metric, and (ii) it partakes in the Newtonian potential $V(r)$. We
reach two key findings. Firstly, the Newtonian limit only emerges
owing to the de Sitter background. Most existing studies of the Newtonian
limit in modified gravity chose to linearize the metric around a locally
flat background. However, this is a \emph{false} vacuum to start with
for pure $\mathcal{R}^{2}$ gravity. These studies unknowingly omitted
the information about $\Lambda$ of the de Sitter background, hence
incapable of attaining a Newtonian behavior in pure $\mathcal{R}^{2}$
gravity. Secondly, as $\Lambda$ appears in $V(r)$ in a \emph{singular}
manner, viz. $V(r)\propto\Lambda^{-1}$, the Newtonian limit for pure
$\mathcal{R}^{2}$ gravity cannot be obtained by any perturbative
approach treating $\Lambda$ as a small parameter.
\end{abstract}
\maketitle

\section{\label{sec:Motivation}Motivation}

The interest in pure $\mathcal{R}^{2}$ gravity has recently experienced
a resurgence. As a member of the $f(\mathcal{R})$ family, it is known
to be ghost-free. This is because when shifting from the Jordan frame
to the Einstein frame, its auxiliary scalar degree of freedom only
involves derivatives of second order and no higher. As a result, the
theory escapes the Ostrogradsky instability that often plagues higher-derivative
theories \citep{Woodard-2007,Woodard-2015}. Unlike other siblings
in the $f(\mathcal{R})$ family, pure $\mathcal{R}^{2}$ gravity enjoys
yet an additional advantage---it is free of any inherent scale. Its
action consists of a single term, $(2\kappa)^{-1}\int d^{4}x\sqrt{-g}\,\mathcal{R}^{2}$,
where the gravitational coupling $\kappa$ is dimensionless. Pure
$\mathcal{R}^{2}$ gravity has a \emph{restricted} scale symmetry
\citep{Edery-2014}, viz. it exhibits invariance under the Weyl transformation,
$g_{\mu\nu}=\Omega^{2}(x)\,\tilde{g}_{\mu\nu}$, with the local conformal
factor $\Omega(x)$ obeying the harmonic condition, $\square\,\Omega=0$.
Theoretical aspects of pure $\mathcal{R}^{2}$ gravity and its implications
for black holes and cosmology have been the subject of active investigation
\citep{Lust-2015-backholes,Clifton-2006,Ferreira-2019,Pravda-2017,Nguyen-2022-Buchdahl,2023-axisym,2023-WH,Alvarez-2018,Rinaldi-2018,Donoghue-2018,Gurses-2012,Frolov-2009,Stelle-1977,Easson-2015,Easson-2017,Murk-2022,Pravda-2018,Stelle-1978,Stelle-2015-a,Stelle-2015-b,Nguyen-2023-Lambda0,Nguyen-2023-Extension,Nguyen-2023-Nontrivial}.\vskip4pt

It is worth noting that the pure $\mathcal{R}^{2}$ action departs
from the conventional Einstein-Hilbert paradigm by excluding the Einstein-Hilbert
term $\frac{c^{4}}{16\pi G}\,\mathcal{R}$. In an important work on
graviton propagators \citep{AlvarezGaume-2015}, Alvarez-Gaume \emph{et
al }discovered that the pure $\mathcal{R}^{2}$ action propagates
a \emph{massless} spin-2 tensor mode \emph{provided that} the background
metric is de Sitter rather than locally flat. This \emph{massless}
mode has a capacity to yield a \emph{long-range} interaction instead
of a short-range Yukawa exchange (the latter would be a typical hallmark
of a \emph{massive} mode, otherwise). Consequently, it would be natural
to expect that the massless spin-2 tensor mode in pure $\mathcal{R}^{2}$
gravity should produce a potential with the correct Newtonian tail
$\sim1/r$, akin to the behavior exhibited by a standard massless
spin-2 tensor mode in General Relativity (GR). The significance of
this mode lies in its potential to confer a proper Newtonian behavior
to the pure $\mathcal{R}^{2}$ theory, despite the \emph{absence}
of the Einstein-Hilbert term in the action. \vskip4pt

For a theory to be a viable description of gravitational phenomena
in the Solar System, it must exhibit a Newtonian limit with the tail
of $1/r$, with $r$ being the distance from a mass source. This limit
was well established in the case of the Einstein-Hilbert action, soon
after the development of GR, through solving the Einstein field equations
in the presence of normal matter. A typical example involves considering
a static point mass $M$ located at the origin, where the stress-energy
tensor takes the form 
\begin{equation}
T_{\mu\nu}=Mc^{2}\,\delta(\vec{r})\,\delta_{\mu}^{0}\,\delta_{\nu}^{0}\label{eq:EMT}
\end{equation}
which has a \emph{non-vanishing} trace, viz. $T:=g^{\mu\nu}T_{\mu\nu}=Mc^{2}\delta(\vec{r})$.
For this stress-energy tensor, the Einstein field equation,
\begin{equation}
\mathcal{R}_{\mu\nu}-\frac{1}{2}g_{\mu\nu}\mathcal{R}=\frac{8\pi G}{c^{4}}\,T_{\mu\nu}\label{eq:GR-eqn}
\end{equation}
is known to admit the Schwarzschild metric as an exterior solution,
given by
\begin{equation}
ds^{2}=-\left(1-\frac{r_{\text{s}}}{r}\right)c^{2}dt^{2}+\left(1-\frac{r_{\text{s}}}{r}\right)^{-1}dr^{2}+r^{2}d\Omega^{2}\label{eq:Schwa}
\end{equation}
where the Schwarzschild radius $r_{\text{s}}$ is equal to $2GM/c^{2}$.
The term $V(r):=-\frac{c^{2}}{2}\frac{r_{\text{s}}}{r}$ represents
a gravitational potential that exhibits the correct Newtonian falloff
$\sim1/r$, with $r_{\text{s}}$ determined by the mass source $M$
and the (Newton) gravitational constant $G$ in the Einstein-Hilbert
action, $\frac{c^{4}}{16\pi G}\int d^{4}x\sqrt{-g}\,\mathcal{R}$.\vskip8pt

Establishing the Newtonian limit becomes intricate when considering
alternative theories of gravity, and one illustrative example is conformal
gravity. The theory is defined by the action $(2\kappa_{\text{CG}})^{-1}\int d^{4}x\sqrt{-g}\ \mathcal{C}_{\mu\nu\rho\sigma}\,\mathcal{C}^{\mu\nu\rho\sigma}$,
where $\mathcal{C}_{\mu\nu\rho\sigma}$ represents the Weyl tensor
and $\kappa_{\text{CG}}$ denotes a dimensionless coupling parameter
\citep{Mannheim-2006}. The field equation of conformal gravity, known
as the Bach equation, is given by \textbf{
\begin{equation}
B_{\mu\nu}:=\left(\nabla^{\rho}\nabla^{\sigma}+\frac{1}{2}\mathcal{R}^{\rho\sigma}\right)\mathcal{C}_{\mu\rho\nu\sigma}=\frac{\kappa_{\text{CG}}}{2}\,T_{\mu\nu}\label{eq:CG-eqn}
\end{equation}
}Interestingly, this equation also allows for a Schwarzschild metric
as a valid solution in vacuum regions where $T_{\mu\nu}=0$. However,
despite this property, establishing the Newtonian limit for conformal
gravity remains an unresolved challenge \citep{Barabash-1999,Barabash-2008}.
One of the key obstacles arises from the \emph{traceless} nature of
the Bach tensor $B_{\mu\nu}$: per Eq.$\,$\eqref{eq:CG-eqn}, the
(conformal) gravitational field can only interact with a stress-energy
tensor that is also traceless, whereas the stress-energy tensor of
ordinary matter, expressed in Eq.$\,$\eqref{eq:EMT}, has a non-zero
trace. Consequently, connecting the Schwarzschild radius $r_{\text{s}}$
and the mass $M$ of the source becomes problematic in conformal gravity.\vskip8pt

In the case of the pure $\mathcal{R}^{2}$ action, the corresponding
field equation is given by
\begin{equation}
\mathcal{R}\Bigl(\mathcal{R}_{\mu\nu}-\frac{1}{4}g_{\mu\nu}\mathcal{R}\Bigr)+\left(g_{\mu\nu}\,\square-\nabla_{\mu}\nabla_{\nu}\right)\mathcal{R}=\frac{\kappa}{2}\,T_{\mu\nu}\label{eq:R2-field-eqn}
\end{equation}
It is known that this equation admits a Schwarzschild metric as a
solution in the exterior region. However, as demonstrated in the case
of conformal gravity, \emph{this still falls short of establishing
the desired Newtonian limit}. To achieve this goal, it is required
to establish a connection between the Schwarzschild radius $r_{\text{s}}$,
the mass $M$, \emph{and} the parameter $\kappa$ of the action $(2\kappa)^{-1}\int d^{4}x\sqrt{-g}\,\mathcal{R}^{2}$.
The challenge lies in the fact that the gravitational coupling $\kappa$
is \emph{dimensionless} and does not resemble the \emph{dimensionful}
Newton constant $G$ that would partake in the regular Newtonian potential,
$V(\vec{r})=-\frac{GM}{r}$. In other words, since the pure $\mathcal{R}^{2}$
theory lacks the (dimensionful) $G$ in its action, one must ``construct''
$G$ from the (dimensionless) $\kappa$. This crucial issue has not
been addressed in the existing literature. The objective of our paper
is to solve Eq.$\,$\eqref{eq:R2-field-eqn} for the stress-energy
tensor expressed in Eq.$\,$\eqref{eq:EMT} and establish the desired
connection between $r_{\text{s}}$ and $\kappa$. The presence of
the Dirac delta function in Eq.$\,$\eqref{eq:EMT} makes the task
delicate. To tackle it, we will introduce a new method utilizing the
Gauss-Ostrogradsky theorem.\vskip8pt

It is important to highlight that previous research on the existence
of a Newtonian limit in pure $\mathcal{R}^{2}$ gravity has predominantly
yielded negative results, as documented in comprehensive studies such
as Refs.$\ $\citep{Capozziello-2009,Capozziello-2010,Stabile-2008}.
These works shared a common characteristic: they employed a weak-field
approximation around a locally flat background. However, as we will
demonstrate in this paper, this practice is not applicable for pure
$\mathcal{R}^{2}$ gravity, which inherently possesses a de Sitter
background. By utilizing a locally flat background, these previous
studies inadvertently \emph{neglected} the crucial information regarding
the scalar curvature of the de Sitter background. As highlighted by
Alvarez-Gaume \emph{et al} in \citep{AlvarezGaume-2015}, it is precisely
the de Sitter background that accommodates the \emph{massless} spin-2
graviton, offering a potential avenue for a \emph{long-range} Newtonian
potential to emerge.\vskip8pt

Our paper is organized as follows: In Section \ref{sec:Warm-up},
we introduce a new method which utilizes the Gauss-Ostrogradsky theorem
and, as a proof of concept, apply it to (re-)derive the Newtonian
potential for the Einstein-Hilbert action. Section \ref{sec:Deriving-for-R2}
encompasses the central outcome of our research, as we employ the
method to solve the field equation of pure $\mathcal{R}^{2}$ gravity,
viz. Eq.$\,$\eqref{eq:R2-field-eqn}, for the stress-energy tensor
given in Eq.$\,$\eqref{eq:EMT}. We present a comprehensive, step-by-step
calculation in full detail, with additional assistance provided in
Appendix \ref{sec:Detailed-calculations}. Section \ref{sec:Why-fail}
delves into the reasons why previous works have been unable to establish
the Newtonian limit for pure $\mathcal{R}^{2}$ gravity. We shed light
on the limitations and shortcomings that hindered their progress in
achieving this goal. Finally, Section \ref{sec:Summary} summarizes
the key steps of our work and offers an outlook.

\section{\label{sec:Warm-up}Re-deriving the Newtonian limit for Einstein-Hilbert
action by way of the divergence theorem}

To solve the complete field equation \eqref{eq:R2-field-eqn}, we
design a specialized method to target the stress-energy tensor $T_{\mu\nu}$
given in \eqref{eq:EMT} which involves the Dirac delta function.
This section serves as a proof-of-concept demonstration, where we
introduce our method and utilize it to re-derive the well-established
Newtonian limit of the standard Einstein-Hilbert theory. In Section
\ref{sec:Deriving-for-R2}, we will apply our method to the pure $\mathcal{R}^{2}$
action.
\noindent \begin{center}
-----------------$\infty$-----------------
\par\end{center}

Consider the following static spherically symmetric line element (with
the speed of light $c$ being explicit):
\begin{align}
ds^{2} & =-f(r)\,c^{2}dt^{2}+\frac{dr^{2}}{f(r)}+r^{2}d\Omega^{2}\label{eq:2.1}\\
d\Omega^{2} & =d\theta^{2}+\sin^{2}\theta d\phi^{2}
\end{align}
For the Einstein-Hilbert action
\begin{equation}
\mathcal{S}_{\text{EH}}=\frac{c^{4}}{16\pi G}\int d^{4}x\sqrt{-g}\,\mathcal{R}\label{eq:EH-action}
\end{equation}
the Einstein field equation is
\begin{equation}
G_{\mu\nu}:=\mathcal{R}_{\mu\nu}-\frac{1}{2}g_{\mu\nu}\mathcal{R}=\frac{8\pi G}{c^{4}}\,T_{\mu\nu}\label{eq:Einstein-tensor}
\end{equation}
with the stress-energy tensor for a static point mass $M$:
\begin{equation}
T_{\mu\nu}=Mc^{2}\,\delta(\vec{r})\,\delta_{\mu}^{0}\,\delta_{\nu}^{0}\label{eq:2.2}
\end{equation}
The 00-component of the Einstein field equation is thus
\begin{equation}
G_{00}=\frac{8\pi GM}{c^{2}}\,\delta(\vec{r})\label{eq:2.3}
\end{equation}
The relevant terms of the Ricci tensor:
\begin{align}
\mathcal{R}_{00} & =f\frac{(r^{2}f')'}{2r^{2}}\label{eq:R00}\\
\mathcal{R} & =\frac{1}{r^{2}}\left[2-(r^{2}f)''\right]\label{eq:Ricci}
\end{align}
give
\begin{align}
G_{00} & =\frac{f}{r^{2}}\left[1-(rf)'\right]\label{eq:G00}
\end{align}
Expressing 
\begin{equation}
f=1+\frac{2}{c^{2}}{\displaystyle V}(r)
\end{equation}
leads to
\begin{align}
G_{00} & =-2\left(1+\frac{2}{c^{2}}V\right)\frac{(rV)'}{c^{2}r^{2}}\label{eq:G00-vs-phi-1}\\
 & \approx-2\frac{(rV)'}{c^{2}r^{2}}\label{eq:G00-vs-phi-2}
\end{align}
in which only the linear term is $V$ is retained.\vskip4pt

To proceed, our strategy is to express $G_{00}$ as a divergence of
a vector field $\vec{A}$ then make use of the Gauss-Ostrogradsky
theorem. Below is how to do so.\vskip4pt

Let us define a vector field $\vec{A}(\vec{r})$ which has only radial
component, i.e.
\begin{equation}
\vec{A}(\vec{r}):=A(r)\,\hat{r}\label{eq:def-A-1}
\end{equation}
with $A(r)$ \emph{judiciously} chosen as 
\begin{equation}
A(r):=\frac{V(r)}{r}\label{eq:def-A-2}
\end{equation}
The divergence of $\vec{A}(\vec{r})$ in the spherical coordinate:
\begin{align}
\vec{\nabla}.\vec{A} & \equiv\frac{1}{r^{2}}\,\frac{\partial}{\partial r}\left(r^{2}A\right)\\
 & =\frac{1}{r^{2}}\,\left(rV(r)\right)'\label{eq:2.4}
\end{align}
Comparing \eqref{eq:G00-vs-phi-2} and \eqref{eq:2.4}, we have
\begin{equation}
\vec{\nabla}.\vec{A}=-\frac{c^{2}}{2}\,G_{00}\label{eq:2.5}
\end{equation}
and, using \eqref{eq:2.3}, a divergence equation
\begin{equation}
\vec{\nabla}.\vec{A}=-4\pi GM\,\delta(\vec{r})\label{eq:2.6}
\end{equation}
Multiply both sides of Eq.$\,$\eqref{eq:2.6} with the 3-volume element
$d^{3}V=r^{2}\,dr\,d\Omega$ and integrate over the ball $V$ of radius
$r$ and center at the origin:
\begin{equation}
\int_{V}d^{3}V\,\bigl(\vec{\nabla}.\vec{A}\bigr)=-4\pi GM\label{eq:2.7}
\end{equation}
Applying the Gauss-Ostrogradsky theorem turns the left-hand-side of
Eq.$\,$\eqref{eq:2.7} into a surface integral on the sphere:
\begin{equation}
\oint_{S}d\vec{S}.\vec{A}=-4\pi GM\label{eq:2.8}
\end{equation}
which, by virtue of spherical symmetry, yields
\begin{equation}
4\pi r^{2}A(r)=-4\pi GM\label{eq:2.9}
\end{equation}
Combining \eqref{eq:2.9} with \eqref{eq:def-A-2}, we readily obtain
\begin{equation}
V(r)=-\frac{GM}{r}\label{eq:2.10}
\end{equation}
which is precisely the Newton law of universal gravitation. This concludes
our proof-of-concept exercise.

\section{\label{sec:Deriving-for-R2}Deriving the Newtonian potential for
pure $\mathcal{R}^{2}$ action}

Let us apply the Gauss-Ostrogradsky procedure designed in the preceding
section to the case at hand
\begin{equation}
\frac{1}{2\kappa}\int d^{4}x\sqrt{-g}\,\mathcal{R}^{2}
\end{equation}
in which $\kappa$ is a \emph{dimensionless} coupling. For brevity,
let us define the following tensor: \footnote{Recall that, for $\frac{1}{2\kappa}f(\mathcal{R})+\mathcal{L}_{\text{m}}$
action, the field equation is: $f'(\mathcal{R})\,\mathcal{R}_{\mu\nu}-\frac{1}{2}f(\mathcal{R})\,g_{\mu\nu}+\left(g_{\mu\nu}\square-\nabla_{\mu}\nabla_{\nu}\right)f'(\mathcal{R})=\kappa\,T_{\mu\nu}$
and the stress-energy tensor: $T_{\mu\nu}:=-\frac{2}{\sqrt{-g}}\frac{\delta(\sqrt{-g}\mathcal{L}_{\text{m}})}{\delta g^{\mu\nu}}.$}
\begin{equation}
X_{\mu\nu}:=\mathcal{R}\Bigl(\mathcal{R}_{\mu\nu}-\frac{1}{4}g_{\mu\nu}\mathcal{R}\Bigr)+\left(g_{\mu\nu}\square-\nabla_{\mu}\nabla_{\nu}\right)\mathcal{R}\label{eq:def-X}
\end{equation}
The pure $\mathcal{R}^{2}$ field equation reads
\begin{equation}
X_{\mu\nu}=\frac{\kappa}{2}\,T_{\mu\nu}\label{eq:field-eqn-R2}
\end{equation}
with the stress-energy tensor given in Eq.$\,$\eqref{eq:2.2}. The
equation to be solved is the 00-component of Eq.$\,$\eqref{eq:field-eqn-R2}:
\begin{equation}
X_{00}=\frac{\kappa}{2}\,Mc^{2}\,\delta(\vec{r})\label{eq:00-field-eqn-R2}
\end{equation}

Our strategy is analogous to the preceding Section: we shall express
$X_{00}$ as the divergence of a (to-be-determined) vector field $\vec{B}(\vec{r})$,
then apply the Gauss-Ostrogradsky theorem to Eq.$\,$\eqref{eq:00-field-eqn-R2}
in order to link $\vec{B}$ with the mass source $M$.\vskip4pt

To proceed, we first need to compute two additional terms in $X_{00}$,
namely, $g_{\mu\nu}\,\square\,\mathcal{R}$ and $\nabla_{\mu}\nabla_{\nu}\mathcal{R}$.
Since $\mathcal{R}$ only depends on $r$, the following expressions
hold: \footnote{Recall that for a scalar field $\phi$: $\square\,\phi=\frac{1}{\sqrt{-g}}\partial_{\mu}(\sqrt{-g}g^{\mu\nu}\partial_{\nu}\phi)$
and $\nabla_{\mu}\nabla_{\nu}\,\phi=\partial_{\mu}\partial_{\nu}\phi-\Gamma_{\mu\nu}^{\lambda}\partial_{\lambda}\phi$.}
\begin{align}
g_{00}\,\square\,\mathcal{R} & =-\frac{f}{\sqrt{-g}}\,\partial_{r}(\sqrt{-g}\,g^{11}\,\partial_{r}\,\mathcal{R})\\
 & =\frac{f}{r^{2}}\,(r^{2}f\mathcal{R}')'\label{eq:3.1}
\end{align}
and (noting that $\partial_{0}\mathcal{R}=0$)
\begin{align}
\nabla_{0}\nabla_{0}\mathcal{R} & =\cancel{\partial_{0}\partial_{0}\mathcal{R}}-\Gamma_{00}^{1}\,\partial_{r}\,\mathcal{R}\\
 & =-\frac{1}{2}\,ff'\mathcal{R}'\label{eq:3.2}
\end{align}
Substituting \eqref{eq:R00}, \eqref{eq:Ricci}, \eqref{eq:3.1},
and \eqref{eq:3.2} into Eq.$\,$\eqref{eq:def-X}:
\begin{align}
X_{00} & =\mathcal{R}\Bigl(\mathcal{R}_{00}-\frac{1}{4}g_{00}\mathcal{R}\Bigr)+\left(g_{00}\,\square-\nabla_{0}\nabla_{0}\right)\mathcal{R}\\
 & =\frac{f}{r^{2}}\biggl[\frac{1}{2}(r^{2}f'\mathcal{R})'+\frac{1}{4}r^{2}\mathcal{R}^{2}-(r^{2}f\mathcal{R}')'\biggr]\label{eq:3.3}
\end{align}
see Eq.$\,$\eqref{eq:X00-simplified} in Appendix \ref{sec:Detailed-calculations}.
It is straightforward to verify that a function $f$ chosen to be
$f=1-\frac{\Lambda}{3}r^{2}$ would result in $\mathcal{R}\equiv4\Lambda$
per Eq.$\,$\eqref{eq:Ricci}, and force $X_{00}$ in Eq.$\,$\eqref{eq:3.3}
to vanish everywhere, thus obeying the vacuo field equation. Let us
then substitute
\begin{equation}
f=1-\frac{\Lambda}{3}r^{2}+\frac{2}{c^{2}}{\displaystyle V(r)}\label{eq:3.4}
\end{equation}
with $V(r)$\emph{ to be determined}, into Eq.$\,$\eqref{eq:3.3}.
We get (see Eq.$\,$\eqref{eq:X00-final} in Appendix \ref{sec:Detailed-calculations}
for step-by-step calculations)
\begin{align}
c^{2}X_{00}(r) & \approx\frac{2}{r^{2}}\Bigl(r^{2}\Bigl(\frac{(r^{2}V)''}{r^{2}}\Bigr)'\Bigr)'\nonumber \\
 & \ \ \ \ \ \ \ -\frac{\Lambda}{r^{2}}\biggl[8(rV)'+\frac{2}{3}\Bigl(r^{5}\Bigl(\frac{(r^{2}V)''}{r^{3}}\Bigr)'\Bigr)'\biggr]\nonumber \\
 & -\frac{2\Lambda}{3}\Bigl(r^{2}\Bigl(\frac{(r^{2}V)''}{r^{2}}\Bigr)'\Bigr)'\nonumber \\
 & \ \ \ \ \ \ \ +\frac{\Lambda^{2}}{3}\biggl[8(rV)'+\frac{2}{3}\Bigl(r^{5}\Bigl(\frac{(r^{2}V)''}{r^{3}}\Bigr)'\Bigr)'\biggr]\label{eq:X00-final-2}
\end{align}
in which only the terms linear in $V$ are retained (i.e., the weak-field
approximation). Note, however, that we retained \emph{all orders}
of $\Lambda$ in Eq.$\,$\eqref{eq:X00-final-2}.\vskip4pt

To find the vector field $\vec{B}(\vec{r})$, we would need to find
the analytical formula for the anti-derivative $\int dr\,r^{2}X_{00}(r)$.
From Eq.$\,$\eqref{eq:X00-final-2}, the anti-derivative can be expressed
as
\begin{equation}
c^{2}\int dr\,r^{2}X_{00}(r):=I(r)-\frac{\Lambda}{3}J(r)\label{eq:3.5}
\end{equation}
in which
\begin{align}
I(r) & :=\int dr\,\biggl\{2\Bigl(r^{2}\Bigl(\frac{(r^{2}V)''}{r^{2}}\Bigr)'\Bigr)'\nonumber \\
 & \ \ \ \ \ \ \ \ -\Lambda\biggl[8(rV)'+\frac{2}{3}\Bigl(r^{5}\Bigl(\frac{(r^{2}V)''}{r^{3}}\Bigr)'\Bigr)'\biggr]\biggr\}\label{eq:3.6}
\end{align}
\begin{align}
J(r) & :=\int dr\,\biggl\{2r^{2}\Bigl(r^{2}\Bigl(\frac{(r^{2}V)''}{r^{2}}\Bigr)'\Bigr)'\nonumber \\
 & \ \ \ -\Lambda\biggl[8r^{2}(rV)'+\frac{2}{3}r^{2}\Bigl(r^{5}\Bigl(\frac{(r^{2}V)''}{r^{3}}\Bigr)'\Bigr)'\biggr]\biggr\}\label{eq:3.7}
\end{align}
The anti-derivative $J(r)$ can be split further into
\begin{align}
J(r) & =2J_{0}(r)-\Lambda\left(8J_{1}(r)+\frac{2}{3}J_{2}(r))\right)\label{eq:def-J}
\end{align}
in which
\begin{align}
J_{0}(r) & :=\int dr\,r^{2}\Bigl(r^{2}\Bigl(\frac{(r^{2}V)''}{r^{2}}\Bigr)'\Bigr)'\label{eq:3.8}\\
J_{1}(r) & :=\int dr\,r^{2}(rV)'\label{eq:3.9}\\
J_{2}(x) & :=\int dr\,r^{2}\Bigl(r^{5}\Bigl(\frac{(r^{2}V)''}{r^{3}}\Bigr)'\Bigr)'\label{eq:3.10}
\end{align}
Full details of manipulation for the anti-derivatives are presented
in Appendix \ref{sec:Detailed-calculations}. We only quote the results
here. Below are Eqs.$\,$\eqref{eq:A.1}, \eqref{eq:A.2}, \eqref{eq:A.3},
and \eqref{eq:A.4} in Appendix \ref{sec:Detailed-calculations}:
\begin{align}
I(r) & =2r^{2}\Bigl(\frac{(r^{2}V)''}{r^{2}}\Bigr)'-\Lambda\biggl[8rV+\frac{2}{3}r^{5}\Bigl(\frac{(r^{2}V)''}{r^{3}}\Bigr)'\biggr]\label{eq:final-I}
\end{align}
\vspace{-.25cm}
\begin{align}
J_{0}(r) & =r^{6}\Bigl(\frac{(r^{2}V)''}{r^{4}}\Bigr)'+6(r^{2}V)'\label{eq:3.11}\\
J_{1}(r) & =r^{3}V-2\int dr\,(r^{2}V)\label{eq:final-J1}\\
J_{2}(x) & =r^{9}\Bigl(\frac{(r^{2}V)''}{r^{5}}\Bigr)'+12r^{4}V'+24\int dr\,(r^{2}V)\label{eq:final-J2}
\end{align}
The latter three expressions, together with Eq.$\,$\eqref{eq:def-J},
yield {[}see Eq.$\,$\eqref{eq:A.5} in Appendix \ref{sec:Detailed-calculations}{]}
\begin{align}
J(r) & =2r^{6}\Bigl(\frac{(r^{2}V)''}{r^{4}}\Bigr)'+12(r^{2}V)'\nonumber \\
 & \ \ \ \ \ -\Lambda\biggl[\frac{2}{3}r^{9}\Bigl(\frac{(r^{2}V)''}{r^{5}}\Bigr)'+8r^{3}(rV)'\biggr]\label{eq:final-J}
\end{align}
From Eqs.$\,$\eqref{eq:final-I} and \eqref{eq:final-J}, we arrive
at a neat expression {[}see Eq.$\,$\eqref{eq:A.6} in Appendix \ref{sec:Detailed-calculations}{]}
\begin{align}
I(r)-\frac{\Lambda}{3}J(r) & =2r^{2}\Bigl(\frac{(r^{2}V)''}{r^{2}}\Bigr)'\nonumber \\
 & \ \ -4\Lambda\biggl[\frac{(r^{4}V)'}{r^{2}}+\frac{1}{3}r^{5}\sqrt{r}\Bigl(\frac{(r^{2}V)''}{r^{3}\sqrt{r}}\Bigr)'\biggr]\nonumber \\
 & \ \ +\frac{\Lambda^{2}}{3}\biggl[\frac{2}{3}r^{9}\Bigl(\frac{(r^{2}V)''}{r^{5}}\Bigr)'+8r^{3}(rV)'\biggr]\label{eq:final-I-J}
\end{align}
Remarkably, the anti-derivative $I(r)-\frac{\Lambda}{3}J(r)$ exists
in \emph{closed-form}. The terms $\int dr\,(r^{2}V)$ in Eqs.$\,$\eqref{eq:final-J1}
and \eqref{eq:final-J2} have managed to cancel themselves out.\vskip4pt

Conversely, Eq.$\,$\eqref{eq:3.5} is equivalent to
\begin{equation}
c^{2}X_{00}(r)=\frac{1}{r^{2}}\frac{d}{dr}\left(I(r)-\frac{\Lambda}{3}J(r)\right)\label{eq:3.12}
\end{equation}
Next, let us define a vector field $\vec{B}(\vec{r})$ which only
has radial component, i.e.
\begin{equation}
\vec{B}(\vec{r}):=B(r)\,\hat{r}\label{eq:def-B-1}
\end{equation}
with $B(r)$ \emph{judiciously} set equal to
\begin{equation}
B(r):=\frac{1}{r^{2}}\left[I(r)-\frac{\Lambda}{3}J(r)\right]\label{eq:def-B-2}
\end{equation}
The divergence of the radial vector field $\vec{B}(\vec{r})$ in the
spherical coordinate is then
\begin{align}
\vec{\nabla}.\vec{B} & \equiv\frac{1}{r^{2}}\,\partial_{r}(r^{2}B(r))\\
 & =\frac{1}{r^{2}}\frac{d}{dr}\left(I(r)-\frac{\Lambda}{3}J(r)\right)\label{eq:3.13}
\end{align}
Comparing Eqs.$\,$\eqref{eq:3.12} and \eqref{eq:3.13}, the $X_{00}$
component is indeed the \emph{divergence} of the vector field $\vec{B}$:
\begin{equation}
\vec{\nabla}.\vec{B}=c^{2}X_{00}\label{eq:3.14}
\end{equation}
thus, by virtue of Eq.$\,$\eqref{eq:00-field-eqn-R2}, giving
\begin{equation}
\vec{\nabla}.\vec{B}=\frac{1}{2}\,\kappa Mc^{4}\,\delta(\vec{r})\label{eq:3.15}
\end{equation}
Multiply both sides of Eq.$\,$\eqref{eq:3.15} with the 3-volume
element $d^{3}V=r^{2}\,dr\,d\Omega$ then integrate over the ball
$V$ of radius $r$ and center at the origin:
\begin{equation}
\int_{V}d^{3}V\,\bigl(\vec{\nabla}.\vec{B}\bigr)=\frac{1}{2}\,\kappa Mc^{4}\label{eq:3.16}
\end{equation}
The Gauss-Ostrogradsky theorem turns the left-hand-side of Eq.$\,$\eqref{eq:3.16}
into a surface integral on the sphere of radius $r$:
\begin{equation}
\oint_{S}d\vec{S}.\vec{B}=\frac{1}{2}\,\kappa Mc^{4}\label{eq:3.17}
\end{equation}
which, by virtue of spherical symmetry, yields
\begin{equation}
4\pi r^{2}B(r)=\frac{1}{2}\,\kappa Mc^{4}\label{eq:3.18}
\end{equation}
Combining Eqs.$\,$\eqref{eq:3.18} with \eqref{eq:def-B-2}, we arrive
at an ODE: 
\begin{align}
I(r)-\frac{\Lambda}{3}J(r) & =\frac{\kappa Mc^{4}}{8\pi}\ \ \ \text{for }\forall r\label{eq:I-J-eqn}
\end{align}
the left-hand-side of which is described by Eq.$\,$\eqref{eq:final-I-J}.
Since $V(r)=A/r$ automatically makes $X_{00}=0$, by substituting
$V(r)=A/r$ into Eq.$\,$\eqref{eq:final-I-J}, we get
\begin{equation}
I(r)-\frac{\Lambda}{3}J(r)=-12\Lambda A\ \ \ \forall r\label{eq:3.19}
\end{equation}
which, together with Eq.$\,$\eqref{eq:I-J-eqn}, produces
\begin{equation}
A=-\frac{\kappa Mc^{4}}{96\pi\Lambda}\label{eq:A-final-R2}
\end{equation}
With $V(r)=A/r$, we thereby obtain the Newtonian law for pure $\mathcal{R}^{2}$
gravity:
\begin{equation}
V(r)=-\frac{\kappa c^{4}}{96\pi\Lambda}\frac{M}{r}\label{eq:Phi-final-R2}
\end{equation}
This concludes our derivation of the SdS metric for pure $\mathcal{R}^{2}$
gravity.

\section{\label{sec:Why-fail}Why have most other existing studies failed
to produce the Newtonian limit in pure $\mathcal{R}^{2}$ gravity?}

The question of whether pure $\mathcal{R}^{2}$ gravity, a theory
that excludes the Einstein-Hilbert term from its action, exhibits
a Newtonian limit has been a contentious topic, although the prevailing
consensus leans towards the negative. It is generally believed that
pure $\mathcal{R}^{2}$ gravity lacks a proper Newtonian limit, as
supported by the comprehensive studies \citep{Capozziello-2009,Capozziello-2010,Stabile-2008},
for example. Consequently, it is commonly accepted that the inclusion
of the Einstein-Hilbert term as a leading term in the action is necessary
to ensure a Newtonian limit, while higher-order terms such as $\mathcal{R}^{2}$
serve as small supplements to the action \citep{Capozziello-2011}.
However, our result expressed in Eq.$\,$\eqref{eq:Phi-final-R2}
strongly challenges these longstanding beliefs.\vskip4pt

To understand why previous studies have fallen short, let us first
review their approaches and identify the shortcomings. Typically,
when addressing this question in a given theory of gravitation, the
common practice was to linearize the metric around the flat background,
$\eta_{\mu\nu}:=(-+++)$, viz.
\begin{equation}
g_{\mu\nu}=\eta_{\mu\nu}+h_{\mu\nu}\label{eq:perturb-flat}
\end{equation}
with $h_{\lyxmathsym{\textmu}\nu}$ treated as a small perturbation.
The essence of the Newtonian limit boils down to determining whether
or not $h_{00}$ satisfies a second-order Poisson equation in the
presence of normal matter. The rationale for employing $\eta_{\mu\nu}$
stemmed from the requirement of asymptotic flatness, which states
that the metric $g_{\mu\nu}$ should approach a Minkowski metric when
far away from all mass sources. Since the expansion around $\eta_{\mu\nu}$,
as expressed in Eq.$\,$\eqref{eq:perturb-flat}, successfully yields
the Newtonian limit in General Relativity (GR), most studies in modified
gravity have adopted this conventional practice without questioning
the domain or limitation of its applicability. However, in the context
of pure $\mathcal{R}^{2}$ gravity, the metric should morph into a
de Sitter cosmic background at large distances, rendering the use
of $\eta_{\mu\nu}$ \emph{inadequate}. \vskip4pt

A prevalent approach in this area has been to establish the Newtonian
limit -- or the lack thereof -- for the full quadratic action:
\begin{equation}
\gamma\,\mathcal{R}+\beta\,\mathcal{R}^{2}+\alpha\,\text{\ensuremath{\mathcal{C}}}^{\mu\nu\rho\sigma}\,\mathcal{C}_{\mu\nu\rho\sigma}\label{eq:full-quadratic-action}
\end{equation}
and subsequently set $\alpha$ and $\gamma$ to zero in the final
analytical result. Among these endeavors, the comprehensive investigation
conducted in Refs.$\ $\citep{Capozziello-2009,Capozziello-2010,Stabile-2008}
can be quoted as a representative example. Their rationale was that
a second-order Poisson equation that is required for the Newtonian
tail, naturally arises from the second-order GR theory but is incompatible
with the fourth-order nature of the $\mathcal{R}^{2}$ theory. Instead,
this latter theory would be expected to support a fourth-order Poisson
equation:
\begin{equation}
\nabla^{4}U(\vec{r})=\delta(\vec{r})\label{eq:4th-Poisson}
\end{equation}
The solution to this equation yields a linear potential, $U(\vec{r})=-\frac{1}{8\pi}r$,
which lacks a desired Newtonian tail $1/r$. \vskip8pt

However, a critical issue emerges. In Ref.$\ $\citep{AlvarezGaume-2015},
Alvarez-Gaume \emph{et al} highlighted an subtle problem with the
full quadratic action. In the absence of matter, the background of
the full action \eqref{eq:full-quadratic-action} (when $\gamma\neq0$)
is Ricci-scalar-flat, i.e., $\mathcal{R}=0$. In contrast, the background
of the pure $\mathcal{R}^{2}$ action (i.e., action \eqref{eq:full-quadratic-action}
with $\gamma=0$) is characterized by a constant Ricci scalar, $\mathcal{R}=4\Lambda$,
where $\Lambda\in\mathbb{R}$. The transition from $\gamma\neq0$
to $\gamma=0$ poses a problem: the former background contains no
information about $\Lambda$ inherent in the latter background. Consequently,
all analytical results obtained for $\gamma\neq0$ cannot be smoothly
extrapolated to the case where $\gamma=0$, as they -- by design
-- lack knowledge about $\Lambda$. In other words, when dealing
with the pure $\mathcal{R}^{2}$ theory, the locally flat metric is
a \emph{false} background to start with. Instead, one must directly
begin with a de Sitter background. Rather than using the conventional
expansion \eqref{eq:perturb-flat}, one must perturb the metric $g_{\lyxmathsym{\textmu}\nu}$
around a de Sitter background, denoted by $\bar{g}_{\mu\nu}$, per
\begin{equation}
g_{\mu\nu}=\bar{g}_{\mu\nu}+h_{\mu\nu}\label{eq:perturb-dS}
\end{equation}

In their work \citep{AlvarezGaume-2015}, Alvarez-Gaume \emph{et al}
employed both expansions \eqref{eq:perturb-flat} and \eqref{eq:perturb-dS}.
They discovered that the pure $\mathcal{R}^{2}$ theory propagates
a \emph{massless} spin-2 tensor mode on a \emph{de Sitter} background
but \emph{not} a flat background. This is a significant result: \emph{a
de Sitter background is essential to support the propagation of massless
spin-2 gravitons, leading to the emergence of the long-range Newtonian
potential.}\vskip4pt

This is precisely where previous studies have gone astray. While using
a standard locally flat metric $\eta_{\mu\nu}$ in the weak-field
expansion is appropriate for GR as well as the quadratic gravity action
\eqref{eq:full-quadratic-action} when $\gamma\neq0$, it becomes
\emph{inadequate} when considering the quadratic gravity action \eqref{eq:full-quadratic-action}
with $\gamma=0$. The vacua of these theories differ, with GR and
quadratic gravity with $\gamma\neq0$ featuring Ricci-scalar flat
vacua ($\mathcal{R}=0$), whereas the quadratic gravity action with
$\gamma=0$ acquires vacua with non-vanishing scalar curvature.\vskip4pt

Aligned with Alvarez-Gaume \emph{et al}'s work, previous studies that
took a similar approach of starting with a de Sitter background have
also observed indications of Newtonian behavior in certain modified
theories of gravity \citep{Nojiri-2004,Clifton-2005,Pereira-2019}.
One particularly interesting study is by Nojiri and Odintsov \citep{Nojiri-2004},
where they intentionally departed from the conventional method and
reported a Newtonian limit for a modified gravity theory that incorporates
$\mathcal{R}^{2}$.

\section{\label{sec:Summary}Summary and outlook}

We have successfully established the existence of a proper Newtonian
limit in pure $\mathcal{R}^{2}$ gravity, a theory that excludes the
Einstein-Hilbert term at the outset:
\begin{equation}
\frac{1}{2\kappa}\int d^{4}x\sqrt{-g}\,\mathcal{R}^{2}\label{eq:7.0-1}
\end{equation}
There were two principal challenges for us to overcome. Firstly, unlike
General Relativity (GR), which is a second-order theory, pure $\mathcal{R}^{2}$
theory is inherently fourth-order, thus naturally resulting in a fourth-order
Poisson equation instead of a second-order Poisson equation that would
be needed for a Newtonian tail. Secondly, the action \eqref{eq:7.0-1}
lacks a Newton constant $G$ to start with; its gravitational coupling
$\kappa$ is \emph{dimensionless} due to scale invariance.\vskip4pt

To derive our result, we solved the pure $\mathcal{R}^{2}$ field
equation
\begin{equation}
\mathcal{R}\Bigl(\mathcal{R}_{\mu\nu}-\frac{1}{4}g_{\mu\nu}\mathcal{R}\Bigr)+\left(g_{\mu\nu}\,\square-\nabla_{\mu}\nabla_{\nu}\right)\mathcal{R}=\frac{\kappa}{2}\,T_{\mu\nu}\label{eq:R2-field-eqn-1}
\end{equation}
\emph{in the presence of a static point mass $M$}, representing normal
matter with the stress-energy tensor  $T_{\mu\nu}=Mc^{2}\,\delta(\vec{r})\,\delta_{\mu}^{0}\,\delta_{\nu}^{0}$.
Guided by Alvarez-Gaume \emph{et al}'s discovery regarding the role
of the de Sitter background, we cast the metric in the form\vspace{-0.2cm}

\small
\begin{equation}
ds^{2}=-\biggl(1-\frac{\Lambda}{3}r^{2}+\frac{2V(r)}{c^{2}}\biggr)\,c^{2}dt^{2}+\frac{dr^{2}}{1-\frac{\Lambda}{3}r^{2}+\frac{2V(r)}{c^{2}}}+r^{2}d\Omega^{2}\label{eq:my-metric-in-disc}
\end{equation}
\normalsize with $V(r)$ being a function to be determined. In the
absence of matter, $V(r)\equiv0$ which results a de Sitter background.\vskip4pt

The treatment of the Dirac delta function in the stress-energy tensor
required special handling. To do so, we introduced a new method via
the Gauss-Ostrogradsky theorem, where we cast the geometric side of
the $00$-component of the field equation as the divergence of an
auxiliary vector field $\vec{B}(\vec{r})$ which is expressible in
terms of the potential $V(r)$. When applying the Gauss-Ostrogradsky
theorem on the divergence term, the field equation, including the
stress-energy tensor of the point mass $M$, is transformed into an
inhomogenous ODE of $V(r)$. From the ODE, the potential $V(r)$ can
be summarily obtained. We illustrated the use of this divergence-based
method for the Einstein-Hilbert action to recover the classic result
$V(r)=-GM/r$, and then went on to apply the method to the pure $\mathcal{R}^{2}$
action. We carried out step-by-step calculations, as detailed in Section
\ref{sec:Deriving-for-R2} and the Appendix.\vskip4pt

For the vacuo outside of the mass source, the function $V(r)$ was
found to have an exact analytical form
\begin{equation}
V(r)=-\frac{\kappa c^{2}}{96\pi\Lambda}\frac{M}{r}\label{eq:V-in-disc}
\end{equation}
The resulting metric \emph{is} Schwarzschild--de Sitter, with the
Schwarzschild radius parameter $r_{\text{s}}$ being fully determined
by the mass source $M$ and the (dimensionless) gravitational coupling
$\kappa$. The term $V(r)$ therefore represents a potential with
the correct Newtonian falloff.\vskip4pt

The parameter $\Lambda$ plays a vital role. It appears in \emph{two}
places: (i) in the background term $-\Lambda r^{2}/3$ and (ii) in
the Newtonian potential $V(r)$. The de Sitter background allows the
emergence of the Newtonian potential naturally. Its scalar curvature
$4\Lambda$ helps transform the \emph{dimensionless} gravitational
coupling $\kappa$ into a \emph{dimensionful} Newton constant per
\begin{equation}
G=\frac{\kappa c^{4}}{96\pi\Lambda}\label{eq:7.1}
\end{equation}
Note that the Newtonian constant $G$ is\emph{ not} a parameter of
the pure $\mathcal{R}^{2}$ action. Rather it is \emph{generated}
from $\kappa$ and $\Lambda$.\vskip4pt

It is evident why our result is beyond the reach for the conventional
approach that linearizes the metric around a locally flat background,
as explained in Section \ref{sec:Why-fail}. Whereas a linearization
around $\eta_{\mu\nu}$ is legitimate for the Ricci-scalar-flat vacua,
such as those in GR, the locally flat metric $\eta_{\mu\nu}$ is a
\emph{false }background to be used in pure $\mathcal{R}^{2}$ gravity.
Previous works that relied on the locally flat background inadvertently
\emph{omitted} crucial information, namely, the curvature $4\Lambda$
of the background metric.\vskip4pt

Moreover, the participation of $\Lambda$ in the Newtonian potential
is \emph{singular}, as it appears in the denominator of expression
\eqref{eq:V-in-disc}. Consequently, perturbative techniques that
treat $\Lambda$ order-by-order cannot achieve this analytical form
of $V(r)$. Our derivation fully incorporates the effects of $\Lambda$
non-perturbatively, hence able to unveil the singular relationship
between the potential $V(r)$ and $\Lambda$. \emph{In contrast, the
conventional method of linearization around $\eta_{\lyxmathsym{\textmu}\nu}$
fails to capture this singular relation}.\vskip4pt

In summary, building upon Alvarez-Gaume et al's discovery of the crucial
role played by the de Sitter background in enabling the existence
of massless spin-2 gravitons \citep{AlvarezGaume-2015}, our study
strengthens their findings by establishing Newtonian behavior for
pure $\mathcal{R}^{2}$ gravity.\vskip8pt

\textbf{\emph{Outlook}}\emph{.}---In a broader context, the capability
of the $\mathcal{R}^{2}$ term alone to produce a Newtonian potential
opens up the possibility of exploring theories that dispense the Einstein-Hilbert
term from the outset. This advancement paves the way for investigating
classically scale invariant theories of gravity that incorporate the
Glashow-Weinberg-Salam model of particle physics, showcasing significant
potential for further exploration \citep{Einhorn-2019,Salvio-2014}.
\begin{acknowledgments}
I thank Timothy Clifton for his constructive correspondences regarding
his joint work with the late John Barrow\linebreak \citep{Clifton-2005}.
I further thank Richard Shurtleff and Tiberiu Harko for their helps
and comments. Maciej Dunajski helpfully pointed out an affinity between
our work and his recent proposal in \citep{Dunajski-2023}, opening
up potential avenues of exploration and enriching the scope of this
study.
\noindent \begin{center}
-----------------$\infty$-----------------
\par\end{center}
\end{acknowledgments}

\appendix

\section{\label{sec:Detailed-calculations}$\,$ DETAILED CALCULATIONS}

In support of Section \ref{sec:Deriving-for-R2}, below is the detail
calculation being carried out step-by-step.\small

\vskip16pt

\noindent \rule[0.5ex]{1\columnwidth}{1pt}

\noindent $ $

In this Appendix we shall suppress the speed of light $c$ for the
sake of clarity. The metric is
\begin{equation}
ds^{2}=-f(r)dt^{2}+\frac{dr^{2}}{f(r)}+r^{2}d\theta^{2}+r^{2}\sin^{2}\theta d\phi^{2}
\end{equation}
in which
\begin{equation}
f(r)=1-\frac{\Lambda}{3}r^{2}+2V(r)\label{eq:def-f-App}
\end{equation}
The terms that are relevant for our calculations in this Appendix
are (with prime denoting derivative with respect to $r$):
\begin{align}
g_{00} & =-f\\
\mathcal{R}_{00} & =f\frac{(r^{2}f')'}{2r^{2}}=f\frac{(rf)''}{2r}\\
\mathcal{R} & =\frac{1}{r^{2}}\left[2-(r^{2}f)''\right]\\
\square\,\mathcal{R} & =\frac{1}{r^{2}}(r^{2}f\mathcal{R}')'\\
\Gamma_{00}^{1} & =\frac{1}{2}\,ff'
\end{align}
giving

\noindent 
\begin{align}
g_{00}\,\square\,\mathcal{R} & =-\frac{f}{r^{2}}(r^{2}f\mathcal{R}')'
\end{align}

\noindent 
\begin{align}
\nabla_{0}\nabla_{0}\,\mathcal{R} & =\cancel{\partial_{0}\partial_{0}\mathcal{R}}-\Gamma_{00}^{1}\,\partial_{r}\,\mathcal{R}\\
 & =-\frac{1}{2}\,ff'\mathcal{R}'
\end{align}
and
\begin{align}
X_{00} & :=\mathcal{R}(\mathcal{R}_{00}-\frac{1}{4}g_{00}\mathcal{R})+(g_{00}\,\square-\nabla_{0}\nabla_{0})\mathcal{R}\\
 & =\mathcal{R}\Bigl(\frac{1}{2r^{2}}f(r^{2}f')'+\frac{1}{4}f\mathcal{R}\Bigr)-\frac{f}{r^{2}}(r^{2}f\mathcal{R}')'+\frac{1}{2}ff'\mathcal{R}'\\
 & =\frac{f}{r^{2}}\Bigl[\frac{1}{2}(r^{2}f'\mathcal{R})'+\frac{1}{4}r^{2}\mathcal{R}^{2}-(r^{2}f\mathcal{R}')'\Bigr]\label{eq:X00-simplified}
\end{align}
\vskip8pt

\noindent \rule[0.5ex]{1\columnwidth}{1pt}

\noindent \vskip12pt

\noindent Plugging $f$ from \eqref{eq:def-f-App} into \eqref{eq:X00-simplified}.
The signs $\approx$ when used below mean that we keep only terms
that are linear in $V$.
\begin{align}
X_{00} & =\frac{f}{r^{2}}\Biggl[\frac{1}{2}\biggl(r^{2}\Bigl(1-\frac{\Lambda}{3}r^{2}+2V\Bigr)'\Bigl(4\Lambda-2\frac{(r^{2}V)''}{r^{2}}\Bigr)\biggr)'\nonumber \\
 & \ \ \ \ \ +\frac{1}{4}r^{2}\Bigl(4\Lambda-2\frac{(r^{2}V)''}{r^{2}}\Bigr)^{2}\nonumber \\
 & \ \ \ \ \ -\biggl(r^{2}\Bigl(1-\frac{\Lambda}{3}r^{2}+2V\Bigr)\Bigl(4\Lambda-2\frac{(r^{2}V)''}{r^{2}}\Bigr)'\biggr)'\Biggr]
\end{align}
\begin{align}
 & \ \ =\frac{f}{r^{2}}\Biggl[\biggl(\Bigl(-\frac{2\Lambda}{3}r^{3}+2r^{2}V'\Bigr)\Bigl(2\Lambda-\frac{(r^{2}V)''}{r^{2}}\Bigr)\biggr)'\nonumber \\
 & \ \ \ \ \ \ \ +r^{2}\biggl(4\Lambda^{2}-4\Lambda\frac{(r^{2}V)''}{r^{2}}+\frac{((r^{2}V)'')^{2}}{r^{4}}\biggr)\nonumber \\
 & \ \ \ \ \ \ \ \ \ \ +2\biggl(r^{2}\Bigl(1-\frac{\Lambda}{3}r^{2}+2V\Bigr)\Bigl(\frac{(r^{2}V)''}{r^{2}}\Bigr)'\biggr)'\Biggr]
\end{align}
\begin{align}
 & \approx\frac{f}{r^{2}}\Biggl[\biggl(-\frac{4\Lambda^{2}}{3}r^{3}+4\Lambda r^{2}V'+\frac{2\Lambda}{3}r(r^{2}V)''\biggr)'\nonumber \\
 & \ \ \ \ \ \ \ \ \ \ \ +4\Lambda^{2}r^{2}-4\Lambda(r^{2}V)''\nonumber \\
 & \ \ \ \ \ \ \ \ \ \ \ \ \ \ \ +2\biggl(r^{2}\Bigl(1-\frac{\Lambda}{3}r^{2}\Bigr)\Bigl(\frac{(r^{2}V)''}{r^{2}}\Bigr)'\biggr)'\Biggr]
\end{align}
\begin{align}
 & \ \ =\frac{f}{r^{2}}\Biggl[-4\Lambda^{2}r^{2}+4\Lambda(r^{2}V')'+\frac{2\Lambda}{3}(r(r^{2}V)'')'\nonumber \\
 & \ \ \ \ \ \ \ \ \ \ \ \ \ +4\Lambda^{2}r^{2}-4\Lambda(r^{2}V)''\nonumber \\
 & \ \ \ \ \ \ \ \ \ +2\Bigl(r^{2}\Bigl(\frac{(r^{2}V)''}{r^{2}}\Bigr)'\Bigr)'-\frac{2\Lambda}{3}\Bigl(r^{4}\Bigl(\frac{(r^{2}V)''}{r^{2}}\Bigr)'\Bigr)'\Biggr]
\end{align}
\begin{align}
 & \ \ \ \ \ =\frac{1}{r^{2}}\Bigl(1-\frac{\Lambda}{3}r^{2}+2V\Bigr)\Biggl\{2\Bigl(r^{2}\Bigl(\frac{(r^{2}V)''}{r^{2}}\Bigr)'\Bigr)'\nonumber \\
 & \ \ \ \ \ \ \ \ \ \ \ \ \ \ \ \ \ -\Lambda\biggl[8(rV)'+\frac{2}{3}\Bigl(r^{5}\Bigl(\frac{(r^{2}V)''}{r^{3}}\Bigr)'\Bigr)'\biggr]\Biggr\}
\end{align}
\begin{align}
 & \ \ \approx\frac{2}{r^{2}}\Bigl(r^{2}\Bigl(\frac{(r^{2}V)''}{r^{2}}\Bigr)'\Bigr)'\nonumber \\
 & \ \ \ \ \ \ \ \ \ \ \ \ \ \ -\frac{\Lambda}{r^{2}}\biggl[8(rV)'+\frac{2}{3}\Bigl(r^{5}\Bigl(\frac{(r^{2}V)''}{r^{3}}\Bigr)'\Bigr)'\biggr]\nonumber \\
 & \ \ \ \ \ \ \ \ -\frac{2\Lambda}{3}\Bigl(r^{2}\Bigl(\frac{(r^{2}V)''}{r^{2}}\Bigr)'\Bigr)'\nonumber \\
 & \ \ \ \ \ \ \ \ \ \ \ \ \ \ \ \ +\frac{\Lambda^{2}}{3}\biggl[8(rV)'+\frac{2}{3}\Bigl(r^{5}\Bigl(\frac{(r^{2}V)''}{r^{3}}\Bigr)'\Bigr)'\biggr]\label{eq:X00-final}
\end{align}
\rule[0.5ex]{1\columnwidth}{1pt}
\begin{align}
I(r) & :=2\int dr\Bigl(r^{2}\Bigl(\frac{(r^{2}V)''}{r^{2}}\Bigr)'\Bigr)'\nonumber \\
 & \ \ \ \ \ \ -\Lambda\int dr\biggl[8(rV)'+\frac{2}{3}\Bigl(r^{5}\Bigl(\frac{(r^{2}V)''}{r^{3}}\Bigr)'\Bigr)'\biggr]\\
 & =2r^{2}\Bigl(\frac{(r^{2}V)''}{r^{2}}\Bigr)'-\Lambda\biggl[8rV+\frac{2}{3}r^{5}\Bigl(\frac{(r^{2}V)''}{r^{3}}\Bigr)'\biggr]\label{eq:A.1}
\end{align}
\rule[0.5ex]{1\columnwidth}{1pt}
\begin{align}
J_{0}(r) & :=\int drr^{2}\Bigl(r^{2}\Bigl(\frac{(r^{2}V)''}{r^{2}}\Bigr)'\Bigr)'\\
 & =\int d\Bigl(r^{2}\Bigl(r^{2}\Bigl(\frac{(r^{2}V)''}{r^{2}}\Bigr)'\Bigr)\Bigr)\nonumber \\
 & \ \ \ \ \ \ \ \ \ \ \ \ -2\int drr\Bigl(r^{2}\Bigl(\frac{(r^{2}V)''}{r^{2}}\Bigr)'\Bigr)
\end{align}
\begin{align}
 & =r^{4}\Bigl(\frac{(r^{2}V)''}{r^{2}}\Bigr)'-2\int drr^{3}\Bigl(\frac{(r^{2}V)''}{r^{2}}\Bigr)'\\
 & =r^{4}\Bigl(\frac{(r^{2}V)''}{r^{2}}\Bigr)'-2\int d\Bigl(r^{3}\Bigl(\frac{(r^{2}V)''}{r^{2}}\Bigr)\Bigr)\nonumber \\
 & \ \ \ \ \ \ \ \ \ \ \ \ \ \ +6\int drr^{2}\Bigl(\frac{(r^{2}V)''}{r^{2}}\Bigr)
\end{align}
\begin{align}
 & =r^{4}\Bigl(\frac{(r^{2}V)''}{r^{2}}\Bigr)'-2r(r^{2}V)''+6(r^{2}V)'\\
 & =r^{6}\Bigl(\frac{(r^{2}V)''}{r^{4}}\Bigr)'+6(r^{2}V)'\label{eq:A.2}
\end{align}
\rule[0.5ex]{1\columnwidth}{1pt}
\begin{align}
J_{1}(r) & :=\int drr^{2}(rV)'\\
 & =\int d\left(r^{2}(rV)\right)-2\int drr(rV)\\
 & =r^{3}V-2\int dr(r^{2}V)\label{eq:A.3}
\end{align}
\rule[0.5ex]{1\columnwidth}{1pt}
\begin{align}
J_{2}(r) & :=\int drr^{2}\Bigl(r^{5}\Bigl(\frac{(r^{2}V)''}{r^{3}}\Bigr)'\Bigr)'\\
 & =\int dr^{2}\Bigl(r^{5}\Bigl(\frac{(r^{2}V)''}{r^{3}}\Bigr)'\Bigr)\nonumber \\
 & \ \ \ \ \ \ \ \ \ \ \ -2\int drr\Bigl(r^{5}\Bigl(\frac{(r^{2}V)''}{r^{3}}\Bigr)'\Bigr)
\end{align}
\begin{align}
 & =r^{7}\Bigl(\frac{(r^{2}V)''}{r^{3}}\Bigr)'-2\int drr^{6}\Bigl(\frac{(r^{2}V)''}{r^{3}}\Bigr)'\\
 & =r^{7}\Bigl(\frac{(r^{2}V)''}{r^{3}}\Bigr)'-2\int d\Bigl(r^{6}\Bigl(\frac{(r^{2}V)''}{r^{3}}\Bigr)\Bigr)\nonumber \\
 & \ \ \ \ \ \ \ \ \ \ \ \ \ \ \ \ \ \ \ \ \ \ \ \ \ \ +12\int drr^{5}\Bigl(\frac{(r^{2}V)''}{r^{3}}\Bigr)
\end{align}
\begin{align}
 & =r^{7}\Bigl(\frac{(r^{2}V)''}{r^{3}}\Bigr)'-2r^{3}(r^{2}V)''+12\int drr^{2}(r^{2}V)''\\
 & =r^{7}\Bigl(\frac{(r^{2}V)''}{r^{3}}\Bigr)'-2r^{3}(r^{2}V)''\nonumber \\
 & \ \ \ \ \ \ \ \ \ \ +12\int d\left(r^{2}(r^{2}V)'\right)-24\int drr(r^{2}V)'
\end{align}
\begin{align}
 & =r^{7}\Bigl(\frac{(r^{2}V)''}{r^{3}}\Bigr)'-2r^{3}(r^{2}V)''+12r^{2}(r^{2}V)'\nonumber \\
 & \ \ \ \ \ \ \ \ \ \ \ \ -24\int dr\left(r(r^{2}V)\right)+24\int dr(r^{2}V)
\end{align}
\begin{align}
 & =r^{7}\Bigl(\frac{(r^{2}V)''}{r^{3}}\Bigr)'-2r^{3}(r^{2}V)''+12r^{2}(r^{2}V)'\nonumber \\
 & \ \ \ \ \ \ \ \ \ \ \ \ \ \ \ \ \ \ \ \ \ \ -24r^{3}V+24\int dr(r^{2}V)\\
 & =r^{9}\Bigl(\frac{(r^{2}V)''}{r^{5}}\Bigr)'+12r^{4}V'+24\int dr(r^{2}V)\label{eq:A.4}
\end{align}
 \rule[0.5ex]{1\columnwidth}{1pt}
\begin{align}
J(r) & :=2J_{0}(r)-\Lambda\Bigl[8J_{1}(r)+\frac{2}{3}J_{2}(r)\Bigr]\\
 & =2\biggl[r^{6}\Bigl(\frac{(r^{2}V)''}{r^{4}}\Bigr)'+6(r^{2}V)'\biggr]\nonumber \\
 & -8\Lambda\left[r^{3}V-2\int dr(r^{2}V)\right]\nonumber \\
 & -\frac{2\Lambda}{3}\biggl[r^{9}\Bigl(\frac{(r^{2}V)''}{r^{5}}\Bigr)'+12r^{4}V'+24\int dr(r^{2}V)\biggr]
\end{align}
\begin{align}
 & =2r^{6}\Bigl(\frac{(r^{2}V)''}{r^{4}}\Bigr)'+12(r^{2}V)'\nonumber \\
 & -\Lambda\biggl[\frac{2}{3}r^{9}\Bigl(\frac{(r^{2}V)''}{r^{5}}\Bigr)'+8r^{4}V'+8r^{3}V\biggr]
\end{align}
\begin{align}
 & =2r^{6}\Bigl(\frac{(r^{2}V)''}{r^{4}}\Bigr)'+12(r^{2}V)'\nonumber \\
 & \ \ \ \ \ \ \ -\Lambda\biggl[\frac{2}{3}r^{9}\Bigl(\frac{(r^{2}V)''}{r^{5}}\Bigr)'+8r^{3}(rV)'\biggr]\label{eq:A.5}
\end{align}
\rule[0.5ex]{1\columnwidth}{1pt}

$ $

\noindent 
\begin{align}
I(r)-\frac{\Lambda}{3}J(r) & =2r^{2}\Bigl(\frac{(r^{2}V)''}{r^{2}}\Bigr)'-\Lambda\biggl[8rV+\frac{2}{3}r^{5}\Bigl(\frac{(r^{2}V)''}{r^{3}}\Bigr)'\biggr]\nonumber \\
 & \ \ -\frac{\Lambda}{3}\Biggl\{2r^{6}\Bigl(\frac{(r^{2}V)''}{r^{4}}\Bigr)'+12(r^{2}V)'\nonumber \\
 & \ \ \ \ \ -\Lambda\biggl[\frac{2}{3}r^{9}\Bigl(\frac{(r^{2}V)''}{r^{5}}\Bigr)'+8r^{3}(rV)'\biggr]\Biggr\}
\end{align}
\begin{align}
 & \ \ \ \ =2r^{2}\Bigl(\frac{(r^{2})''}{r^{2}}\Bigr)'-\Lambda\Biggl[8rV+\frac{2}{3}r^{5}\Bigl(\frac{(r^{2}V)''}{r^{3}}\Bigr)'\nonumber \\
 & \ \ \ \ \ \ \ \ \ \ \ \ \ \ \ \ \ \ \ \ \ \ \ \ \ \ \ \ \ +\frac{2}{3}r^{6}\Bigl(\frac{(r^{2}V)''}{r^{4}}\Bigr)'+4(r^{2}V)'\Biggr]\nonumber \\
 & \ \ \ \ \ \ \ \ \ \ \ \ \ \ \ \ \ +\frac{\Lambda^{2}}{3}\biggl[\frac{2}{3}r^{9}\Bigl(\frac{(r^{2}V)''}{r^{5}}\Bigr)'+8r^{3}(rV)'\biggr]
\end{align}
\begin{align}
 & \ \ \ \ =2r^{2}\Bigl(\frac{(r^{2}V)''}{r^{2}}\Bigr)'-\Lambda\Biggl[8rV+\frac{2}{3}r^{5}\Bigl(\frac{(r^{2}V)''}{r^{3}}\Bigr)'\nonumber \\
 & \ \ \ \ \ \ \ \ \!\ \ \ \ \ \ \ \ \ \ \ \ \ \ \ \ \ \ \ \ \ \ +\frac{2}{3}r^{6}\Bigl(\frac{(r^{2}V)''}{r^{4}}\Bigr)'+4(r^{2}V)'\Biggr]\nonumber \\
 & \ \ \ \ \ \ \ \ \ \ \ \ \ \ \ \ +\frac{\Lambda^{2}}{3}\biggl[\frac{2}{3}r^{9}\Bigl(\frac{(r^{2}V)''}{r^{5}}\Bigr)'+8r^{3}(rV)'\biggr]
\end{align}
\begin{align}
 & \ \ \ \ =2r^{2}\Bigl(\frac{(r^{2}V)''}{r^{2}}\Bigr)'\nonumber \\
 & \ \ \ \ \ \ \ \ \ \ \ \ -4\Lambda\biggl[\frac{(r^{4}V)'}{r^{2}}+\frac{1}{3}r^{5}\sqrt{r}\Bigl(\frac{(r^{2}V)''}{r^{3}\sqrt{r}}\Bigr)'\biggr]\nonumber \\
 & \ \ \ \ \ \ \ \ \ \ \ \ \ \ \ \ +\frac{\Lambda^{2}}{3}\biggl[\frac{2}{3}r^{9}\Bigl(\frac{(r^{2}V)''}{r^{5}}\Bigr)'+8r^{3}(rV)'\biggr]\label{eq:A.6}
\end{align}

\end{document}